\documentclass{iopart}
\usepackage{graphicx}
\usepackage{amsfonts}
\usepackage{amssymb}
\usepackage{mathrsfs}
\usepackage{amsthm}
\usepackage{mathrsfs}
\usepackage{eufrak}
\usepackage[mathscr]{eucal}

\begin{document}

\newtheorem{twr}{Theorem}
\newtheorem{lem}{Lemma}
\newtheorem{df}{Definition}
\newtheorem{ex}{Example}
\newtheorem{pr}{Proposition}

\title{Spectral properties of entanglement witnesses}
\author{G Sarbicki}
\address{Insitute of Physics, Nicolaus Copernicus University \\
Grudziadzka 5, 87--100 Toru\'n, Poland}
\ead{gniewko@fizyka.umk.pl}

\begin{abstract}
Entanglement witnesses are observables which when measured, detect entanglement in a measured composed system. It is shown  what  kind of relations between eigenvectors of an observable should be fulfilled, to allow an observable to be an entanglement witness. Some restrictions on the signature of entaglement witnesses, based on an algebraic-geometrical theorem will be given. The set of entanglement witnesses is linearly isomorphic to the set of maps between matrix algebras which are positive, but not completely positive. A translation of the results to the language of positive maps is also given. The properties of entanglement witnesses and positive maps express as special cases of general theorems for $k$-Schmidt witnesses and $k$-positive maps. The results are therefore presented in a general framework.
\end{abstract}

\section{Introduction}
The crucial resource in quantum-informational science is quantum entanglement. It provides essential difference between quantum and classical information theory. It is therefore of special importance, to know the techniques of measuring and quantyfying entanglement. Despite of enormous efforts in the last decade, and a number of significant partial results, the problem how to determine that a given density matrix possesses this desired resource or not, is not solved in general. One of the most important methods of detecting entanglement exploits a class of special observables called \textit{entanglement witnesses}. In this paper the entanglement witnesses are considered in the language of their spectral decomposition. 

Only finite-level quantum systems, which are composed of two subsystems, will be considered. The first one is a $d_1$-level system, and the second one is a $d_2$-level system. We assume, that $d_1 \le d_2$. The Hilbert space of the system is then a tensor product of the Hilbert spaces of its subsystems: $\mathcal{H} = \mathbb{C}^{d_1} \otimes \mathbb{C}^{d_2}$. By fixing the orthogonal basis $\{ e_{i} \}_{i=1}^{d_1}$ as the basis of $\mathbb{C}^{d_1}$, and the orthogonal basis $\{ f_{j} \}_{j=1}^{d_2}$ as the basis of $\mathbb{C}^{d_2}$, one gets the linear isomorphism
\begin{displaymath}
\mathfrak{A}: \mathcal{H} \to \mathbb{M}_{\mathbb{C}}(d_1 \times d_2)
\end{displaymath}
between vectors in $\mathcal{H} = \mathbb{C}^{d_1} \otimes \mathbb{C}^{d_2}$ and $d_1 \times d_2$ complex matrices of their coordinates, defined by the formula
\begin{equation}
[\mathfrak{A}(\Psi)]_{i,j} =  \langle e_{i} \otimes f_{j} | \Psi \rangle
\label{mathfrakA}
\end{equation}
The vector $\Psi$ can be therefore expressed by the matrix $\mathfrak{A}(\Psi)$ as
\begin{displaymath}
 \Psi = \sum_{i=1}^{d_1} \sum_{j=1}^{d_2} [\mathfrak{A}(\Psi)]_{ij} | i \rangle | j \rangle
\end{displaymath}

Now the \textit{Schmidt rank} of a vector $\Psi$ in $\mathcal{H}$ is defined as the rank of its coordinate matrix $\mathfrak{A}(\Psi)$. Denote the set of vectors of the Schmidt rank non-greater than $k$ as $\mathcal{S}_{k}$. We then have the ascending sequence of closures of orbits of the group $GL(d_1) \times GL(d_2)$
\begin{displaymath}
\mathcal{S}_{1} \subsetneq \dots \subsetneq \mathcal{S}_{d_1} = \mathbb{C}^{d_1} \otimes \mathbb{C}^{d_2}.
\end{displaymath}
Any matrix $A \in \mathbb{M}_{\mathbb{C}}(d_1 \times d_2)$ of rank one can be written as $v^{T}u$, where $v$ and $u$ are vectors respectively in $\mathbb{C}^{d_1}$ and $\mathbb{C}^{d_2}$. Taking $a \in \mathbb{C}^{*}$ one gets the new pair $u' = a \cdot u$, $v'=v/a$ representing the same vector. Aliasing such pairs, we get the isomorphism $\mathbb{C}^{d_1} \times \mathbb{C}^{d_2} / \mathbb{C} \cong \mathcal{S}_{1}$.

Next define a projection
\begin{displaymath}
\mathfrak{P}:|\Psi \rangle \to \frac{|\Psi \rangle \langle \Psi |}{\langle \Psi | \Psi \rangle}
\end{displaymath}
from $\mathbb{C}^{d_1} \otimes \mathbb{C}^{d_2}$ to the projective space $\mathbb{C}P^{d_1 \times d_2 - 1}$. The projective space $\mathbb{C}P^{d_1 \times d_2 - 1} = \mathbb{C}^{d_1} \otimes \mathbb{C}^{d_2} / \mathbb{C}$ is the space of rays in $\mathbb{C}^{d_1} \otimes \mathbb{C}^{d_2}$, i.e. the space $\mathbb{C}^{d_1} \otimes \mathbb{C}^{d_2} \setminus \{ 0 \}$ where the vectors of the same direction (equal up to multiplication by a non-zero scalar) are aliased (for a detailed discusion about projective geometry see \cite{Cox}, \cite{Hartshorne}). The space of rays of the Hilbert space of a given quantum system is in one-to-one correspondence with the set of pure quantum states of the system.

Now projecting the above sequence of inclusions, one gets the following new ascending sequence of sets of projectors onto one-dimensional subspaces
\vspace{0.2cm}
\begin{displaymath}
\begin{array}{ccccccccc}
& & \mathcal{S}_{1} & \subsetneq & \dots & \subsetneq & \mathcal{S}_{d_1} & = & \mathbb{C}^{d_1} \otimes \mathbb{C}^{d_2}
\\ \\
& & \downarrow \mathfrak{P} & & & & \downarrow \mathfrak{P} & & \downarrow \mathfrak{P} \\ \\
\mathbb{C}P^{d_1-1} \times \mathbb{C}P^{d_2-1} & \cong & 
\mathfrak{P}(\mathcal{S}_{1}) & \subsetneq & \dots & \subsetneq & \mathfrak{P}(\mathcal{S}_{d_1}) & = & \mathbb{C}P^{d_1 \times d_2-1}
\end{array},
\end{displaymath}
\vspace{0.2cm}
where the isomorphism in the lower sequence constitutes the Segre embeding $\mathbb{C}P^{d_1-1} \times \mathbb{C}P^{d_2-1} \hookrightarrow \mathbb{C}P^{d_1 \times d_2-1}$ (see \cite{Cox}, \cite{Hartshorne} for details). The elements of the set $\mathfrak{P}(\mathcal{S}_{1})$ are called \textit{pure separable states}. We define \textit{mixed separable states} in accord with the work of Werner \cite{Wern} as convex combinations of pure separable states.

Following Terhal et al. \cite{TerhHor}, we extend this definition to higher $k$ and define \textit{density operators of Schmidt number k} (\cite{TerhHor}, \cite{Sanp}):
\begin{df}
 The state $\rho$ is called to be of the Schmidt number $k$, when it can be decomposed as a convex combination of projectors
\begin{displaymath}
 \rho = \sum_{i} p_{i} | \psi_{i} \rangle \langle \psi_{i} |,
\end{displaymath}
 where the Schmidt ranks of the vectors $\psi_{i}$ is less or equal $k$, and cannot be decomposed as a convex sum of projectors onto vectors of Schmidt rank less than $k$.
\end{df}

The set of all states of Schmidt number less or equal to $k$ is convex --- it is the convex roof of the set $\mathfrak{P}(\mathcal{S}_{k})$.
The problem of separability is then a problem of membership in a convex set, whith given extremal points.
One of the ways to handle this problem is by using the concept of \textit{entanglement witnesses} introduced in \cite{Terh1}. Taking any entangled state $\rho$, we have two compact, convex, non-empty subsets of the linear coset of operators of rank $1$ - the set of separable states and the singleton of the chosen entangled state. Now according to Banach separation theorem, there exists an affine subspace $\hat{V}$ of codimension $1$ (in the considered coset), which separates these two sets. We can extend now this subspace uniquely to a linear subspace $V$ of codimension $1$ in the Hilbert space of Hermitian operators. Now using the self duality of this Hilbert space, one can assign to the subspace $V$ of codimension $1$ the unique (up to multiplication by non-zero scalar) observable $W$, such that $V$ is its orthogonal complement.
Such an observable is called \textit{entanglement witness}. We easly extend this definition to $k$-Schmidt witnesses (see \cite{Sanp}).

\begin{df}The $k$-Schmidt witness is an observable which fulfils the conditions:
\begin{itemize}
\item $\forall \Psi \in \mathcal{S}_{k} \ \langle \Psi | W | \Psi \rangle \ge 0$ (all states of Schmidt number less or equal to $k$ are on the same side of $V$)
\item for some $\rho \quad \langle \rho | W \rangle_{HS} = \mathrm{Tr}(\rho W)< 0$ (the singleton of $\rho$ lies on the other side of $V$)
\end{itemize}
\label{k_ent_wit}
\end{df}

The problem of classification of $k$-Schmidt witnesses remains unsolved in general. In low dimensions ($2 \times 2$, $2 \times 3$) we have such a classification. Any entanglement witness ($1$-Schmidt witness) is of the form
\begin{displaymath}
W = A^{\Gamma} + B, \ \ A, B \ge 0,
\end{displaymath}
where $\Gamma$ denotes the partial transposition in one of the subsystems (see \cite{decomp}). Such witnesses are called \textit{decomposable}, and states which can be detected by witnesses from this class are called NPT (\textit{negative partial transposition}) entangled states. In higher dimensions this class of entanglement witnesses are a proper subset of the set of all witnesses. Entangled states not detected by this class are called PPT (\textit{positive partial transposition}) entangled states. We have a simple citerion to check, whether a given state is NPT. The most interesting are then non-decomposable witnesses and tools to detect PPT entanglement based on them.

\section{The main theorems}

Having a Hermitian observable $W$, we define a decomposition of its domain due to the spectral decomposition of $W$
\begin{displaymath}
\mathbb{C}^{d_1} \times \mathbb{C}^{d_2} = V_{+} \oplus V_{-} \oplus V_{0}
\end{displaymath}
where the \textit{positive subspace} $V_{+}$ is spaned by eigenvectors corresponding to positive eigenvalues, the \textit{negative subspace} $V_{-}$ is spaned by eigenvectors corresponding to negative eigenvalues, and $V_{0}$ is a kernel of $W$. Next due to the spectral decomposition one can represent $W$ as a difference of two positive operators $W = W_{+} - W_{-}$, the first one supported on $V_{+}$, and the second one on $V_{-}$. Their kernels are $\mathrm{ker} W_{\pm} = V_{0} \oplus V_{\mp}$.

Having a vector $\Psi$ given by it's Schmidt decomposition $\Psi = \sum_{i} \lambda_{i} \alpha_{i} \otimes \beta_{i}$, we define a subspace
\begin{displaymath}
\tilde{V}_{\Psi} = \mathrm{span} \{ \ \mathrm{span} \{ \alpha_{i} \} \otimes \mathbb{C}^{d_2} \ \cup \ \mathbb{C}^{d_1} \otimes \mathrm{span} \{ \beta_{i} \} \ \}.
\end{displaymath}

It is also possible to define this subspace independently of the basis as
\begin{displaymath}
\tilde{V}_{\Psi} = \mathrm{span} \{ \ \mathrm{Im} \mathrm{Tr}_{2} | \Psi \rangle \langle \Psi | \otimes \mathbb{C}^{d_2} \ \cup \ \mathbb{C}^{d_1} \otimes \mathrm{Im} \mathrm{Tr}_{1} | \Psi \rangle \langle \Psi | \ \}
\end{displaymath}
Here $\mathrm{Im}$ denotes the image or range of the operator. Partial traces $\mathrm{Tr}_{1}$ and $\mathrm{Tr}_{2}$ are defined as follows: $\mathrm{Tr}_{1} \rho = \sum_{ik} \delta_{ik} \rho_{ij,kl}$, respectively $\mathrm{Tr}_{2} \rho = \sum_{jl} \delta_{jl} \rho_{ij,kl}$, where $\rho_{ij,kl}$ denotes the matrix element of $\rho$ in standart basis, i.e. $\langle e_i \otimes f_j | \rho | e_k \otimes f_l \rangle$.

\begin{twr} Let $W$ be a $k$-Schmidt witness. Its eigenvectors satisfy the three following conditions:
\begin{enumerate}
\item $V_{-} \supsetneq \{ 0 \}$
\item $\forall \Psi \in \mathcal{S}_{k}, \ \Psi \in V_{0} \oplus V_{-} \Rightarrow \ \Psi \in V_{0}$.
\item $\forall \Psi \in \mathcal{S}_{k}, \cap V_{0} \ \ \tilde{V}_{\Psi} \cap (V_{-} \oplus V_{0}) \subset V_{0}$.
\end{enumerate}
\label{main_th_need}
\end{twr}

\textbf{Proof:}
(i) A $k-$ entanglement witness should detect something, so there are some states, for which the mean value of the witness on them is negative (the second condition in the definition \ref{k_ent_wit}). 

(ii) If $\Psi \in \mathcal{S}_{k} \cap V_{0} \oplus V_{-}$, then $\Psi$ is a combination of eigenvectors only from the kernel and $V_{-}$. Due to the first condition in Definion \ref{k_ent_wit}, the mean value of the observable on the vector $\Psi$ should be nonnegative, so $\Psi$ can be only a combination of the eigenvectors from the kernel.

(iii) To prove the neccesity of the third condition, observe that when an observable $W$ is a $k$-SW, then $\forall \Psi \in \mathcal{S}_{k} \ \langle \Psi | W_{+} | \Psi \rangle - \langle \Psi | W_{+} | \Psi \rangle \ge 0$. This implies the condition that the supremum of the function
\begin{displaymath}
F: \mathcal{S}_{k} \ \backslash \ V_{0} \oplus V_{-} \to \mathbb{R} \qquad F(\Psi) = \frac{\langle \Psi | W_{-} | \Psi \rangle}{\langle \Psi | W_{+} | \Psi \rangle}
\end{displaymath}
is less than $1$. The function $F$ plays a key role in this proof.

Let's take a vector $\sum_{i=1}^{k} x_{i} \otimes y_{i} \in \mathcal{S}_{k}$
in $V_{0}$ and some other vector $\sum_{i=1}^{k} \tilde{x}_{i} \otimes \tilde{y}_{i} \in \mathcal{S}_{k}$
. Consider now the family of vectors form $\mathcal{S}_{k}$:
\begin{equation}
\Phi(t) = \sum_{i=1}^{k} (x_{i}+t \tilde{x}_{i}) \otimes (y_{i}+t \tilde{y}_{i})
\label{family}
\end{equation}
This family forms an algebraic curve (over $\mathbb{R}$) of degree $2$. The intersection $\{ \Phi(t) : t \in \mathbb{R} \} \cap V_{0}$ can be one of the folowing sets:
\begin{enumerate}
\item The whole curve - $\{ \Phi(t) : t \in \mathbb{R} \} \subset V_{0}$.
\item $\{ \Phi(t) : t \in \mathbb{R} \}$ meets the kernel in no more than two points. Because $\sum_{i=1}^{k} x_{i} \otimes y_{i} \in V_{0}$, one of these points is $0$.
\end{enumerate}

To begin with, consider the second case. 
In this case the root of the denominator of $F_{\Phi}$ in $0$ is separated, i.e.
\begin{displaymath}
\exists \epsilon: \{ \Phi(t) : t \in (-\epsilon, \epsilon) \} \cap V_{0} = \{ \Phi(0) \} = \sum_{i=1}^{k} x_{i} \otimes y_{i} \in V_{0}.
\end{displaymath}
The restriction of the function $F$ to the set $\Phi((-\epsilon, \epsilon) \setminus  \{ 0 \} )$ gives a fuction $F_{\Phi}: (-\epsilon, \epsilon) \setminus  \{ 0 \} \to \mathbb{R}$, given by the formula
\begin{equation}
F_{\Phi}(t) = \frac{\langle \Phi(t) | W_{-} | \Phi(t) \rangle}{\langle \Phi(t) | W_{+} | \Phi(t) \rangle}
\label{restriction}
\end{equation}

Computing the numerator of $F_{\Phi}$ and taking into account that $W_{\pm} |\Phi \rangle = 0$ one gets
\begin{eqnarray}
\langle \Phi(t) | W_{-} | \Phi(t) \rangle = 
\nonumber \\
\langle \sum_{i=1}^{k} (x_{i} + t \tilde{x}_{i}) \otimes (y_{i} + t \tilde{y}_{i}) | W_{-} | \sum_{i=1}^{k} (x_{i} + t \tilde{x}_{i}) \otimes (y_{i} + t \tilde{y}_{i}) \rangle 
\nonumber \\
= t^{2} (
t^{2} \langle \sum_{i=1}^{k} \tilde{x}_{i} \otimes \tilde{y}_{i} | W_{-} | \sum_{i=1}^{k} \tilde{x}_{i} \otimes \tilde{y}_{i} \rangle 
\nonumber \\
+ 2 t \mathrm{Re} \langle \sum_{i=1}^{k} x_{i} \otimes \tilde{y}_{i} + \sum_{i=1}^{k} \tilde{x}_{i} \otimes y_{i} | W_{-} | \sum_{i=1}^{k} \tilde{x}_{i} \otimes \tilde{y}_{i} \rangle 
\nonumber \\
+ \langle \sum_{i=1}^{k} x_{i} \otimes \tilde{y}_{i} + \sum_{i=1}^{k} \tilde{x}_{i} \otimes y_{i} | W_{-} | \sum_{i=1}^{k} x_{i} \otimes \tilde{y}_{i} + \sum_{i=1}^{k} \tilde{x}_{i} \otimes y_{i} \rangle ).
\label{licznik}
\end{eqnarray}
For the denominator we get
\begin{eqnarray}
\langle \Phi(t) | W_{+} | \Phi(t) \rangle = 
\nonumber \\
\langle \sum_{i=1}^{k} (x_{i} + t \tilde{x}_{i}) \otimes (y_{i} + t \tilde{y}_{i}) | W_{+} | \sum_{i=1}^{k} (x_{i} + t \tilde{x}_{i}) \otimes (y_{i} + t \tilde{y}_{i}) \rangle 
\nonumber \\
= t^{2} (
t^{2} \langle \sum_{i=1}^{k} \tilde{x}_{i} \otimes \tilde{y}_{i} | W_{+} | \sum_{i=1}^{k} \tilde{x}_{i} \otimes \tilde{y}_{i} \rangle 
\nonumber \\
+ 2 t \mathrm{Re} \langle \sum_{i=1}^{k} x_{i} \otimes \tilde{y}_{i} + \sum_{i=1}^{k} \tilde{x}_{i} \otimes y_{i} | W_{+} | \sum_{i=1}^{k} \tilde{x}_{i} \otimes \tilde{y}_{i} \rangle 
\nonumber \\
+ \langle \sum_{i=1}^{k} x_{i} \otimes \tilde{y}_{i} + \sum_{i=1}^{k} \tilde{x}_{i} \otimes y_{i} | W_{+} | \sum_{i=1}^{k} x_{i} \otimes \tilde{y}_{i} + \sum_{i=1}^{k} \tilde{x}_{i} \otimes y_{i} \rangle ).
\label{mianownik}
\end{eqnarray}
For a given $\{ \tilde{x}_{i}, \tilde{y}_{i} \}$, we get a rational function $F_{\Phi}: \mathbb{R} \to \mathbb{R}$. Such a function is the quotient of two quadratic polynomials
\begin{equation}
F_{\{ \tilde{x}_{i}, \tilde{y}_{i} \} }(t) = \frac{at^2 + bt + c}{dt^2 + et + f}.
\label{f_wym}
\end{equation}

Now the assumtion that the supremum of $F$ on its domain is finite, implies that for any curve $\Phi$, the limit of $F_{\Phi}$ in $0$ is finite.

First consider the degenerated subcase, when $\sum_{i=1}^{k} \tilde{x}_{i} \otimes \tilde{y}_{i} \in V_{0}$. Then $a=d=b=e=0$, and $f$ is non-zero (because we assume at the moment, that the curve $\Phi$ is not a subset of $V_{0}$). The limit of $F_{\Phi}$ in $0$ is finite.

We can then consider the generic case, when $d \ne 0$. At the beginning observe, that $f=0 \Rightarrow e=0$, $c=0 \Rightarrow b=0$. Suppose that the denominator has a root in $0$ (by the above observation, it is then a double root), which means that $f=0 \ \iff \ \sum_{i=1}^{k} x_{i} \otimes \tilde{y}_{i} + \sum_{i=1}^{k} \tilde{x}_{i} \otimes y_{i} \in \mathrm{ker} W_{+}$ (by the formula (\ref{mianownik}) and the positivity of $W_{+}$). Finiteness of the limit of $F_{\Phi}$ in $0$ (according to the formula (\ref{licznik}) and the postivity of $W_{-}$) now implies that also $c=0 \ \iff \ \sum_{i=1}^{k} x_{i} \otimes \tilde{y}_{i} + \sum_{i=1}^{k} \tilde{x}_{i} \otimes y_{i} \in \mathrm{ker} W_{-}$ (then also $b=0$ and the numerator has also a double root in $0$). The limit is then equal $a/d$, which is finite. Using the definition of the subspace $\tilde{V}_{\sum x_{i} \otimes y_{i}}$ and considering that $\ker W_{+} \cap W_{-} = V_{0}$, we can rewrite this implication in the following form
\begin{equation}
\forall \phi \in \tilde{V}_{\sum x_{i} \otimes y_{i}} \ \phi \in V_{0} \oplus V_{-} \Rightarrow \phi \in V_{0}.
\label{implikacja}
\end{equation}
what gives the assertion of the theorem.

It remains to consider the situation, when $\{ \Phi(t): t \in \mathbb{R} \} \subset \mathrm{ker} W_{+}$. This implies that
\begin{eqnarray*}
\sum_{i=1}^{k} \tilde{x} \otimes \tilde{y} \in \mathrm{ker} W_{+} \\
\sum_{i=1}^{k} (\tilde{x} \otimes y + x \otimes \tilde{y}) \in \mathrm{ker} W_{+}.
\end{eqnarray*}
But now by the second point of the theorem, also $\{ \Phi(t): t \in \mathbb{R} \} \subset \mathrm{ker} W_{-}$, so in particular
\begin{displaymath}
\sum_{i=1}^{k} (\tilde{x} \otimes y + x \otimes \tilde{y}) \in \mathrm{ker} W_{-}.
\end{displaymath}
Again, identyfying the kernels and using the defnition of $\tilde{V}_{\sum x_{i} \otimes y_{i}}$ we get the assertion of the theorem.
$\square$

Under some stronger assumptions about the subspaces $V_{0}, V_{+}$ and $V_{-}$, one can prove the opposite theorem:

\begin{twr} Consider the decomposition of $\mathbb{C}^{d_1} \otimes \mathbb{C}^{d_2}$ into direct sum of mutual orthogonal subspaces: $\mathbb{C}^{d_1} \otimes \mathbb{C}^{d_2} = V_{+} \oplus V_{0} \oplus V_{-}$. Let $W_{\pm}$ be arbitrary positive operators supported on $V_{\pm}$.

Define $\hat{V} = \bigcap_{\Psi \in \mathcal{S}_{k} \cap V_{0}} \tilde{V}_{\Psi}^{\perp}$. If
\begin{enumerate}
\item $V_{-} \supsetneq \{ 0 \}$
\item $\forall \Psi \in \mathcal{S}_{k}, \ \Psi \in V_{0} \oplus V_{-} \Rightarrow \ \Psi \in V_{0}$
\item $V_{-} \subset \hat{V}$
\end{enumerate}
then for large enough $\lambda$ the observable $\lambda W_{+} - W_{-}$ is an $k$-SW.
\label{suff}
\end{twr}

\textbf{Proof:} 
(i) The first condition guarantees the existence of detected states, so the first condition in the definition (\ref{k_ent_wit}) is fulfilled.

One can decompose any $k-$separable vector $\Psi$ as
\begin{displaymath}
\Psi = \hat{\Psi} + \tilde{\Psi},
\end{displaymath}
where $\hat{\Psi} \in \hat{V}$ and $\tilde{\Psi} \in \hat{V}^{\perp}$. Now using the decomposition $W = W_{+} - W_{-}$ we can write the second condition in the definition (\ref{k_ent_wit}) as
\begin{eqnarray}
\forall \hat{\Psi}  \in \hat{V} \quad \forall \tilde{\Psi} \in \hat{V}^{\perp}: \quad \hat{\Psi} + \tilde{\Psi} \in \mathcal{S}_{k} \ \langle \hat{\Psi} + \tilde{\Psi}| W_{+} | \hat{\Psi} + \tilde{\Psi} \rangle
\nonumber \\
\ge \langle \hat{\Psi} + \tilde{\Psi}| W_{-} | \hat{\Psi} + \tilde{\Psi} \rangle = \langle \hat{\Psi}| W_{-} | \hat{\Psi} \rangle.
\label{warunek}
\end{eqnarray}
The last equality holds because of the third assumption.
When $\hat{\Psi}=0$, then condition (\ref{warunek}) is fulfilled. It is then enough to focus on vectors $\Psi$ for which $\hat{\Psi} \ne 0$. When the condition (\ref{warunek}) is fulfilled for a vector $\Psi$, then also for a vector $\alpha \Psi$, where $\alpha \in \mathbb{C}^{*}$. It is then sufficient to restrict the quantificated set to the set of vectors $\Psi$, for which $||\hat{\Psi}||=1$. Condition (\ref{warunek}) now takes the form
\begin{eqnarray}
\forall \hat{\Psi} \in \hat{V}: ||\hat{\Psi}||=1 \ \forall \tilde{\Psi} \in \hat{V}^{\perp}: \ \hat{\Psi} + \tilde{\Psi} \in \mathcal{S}_{k} \nonumber \\
\langle \hat{\Psi} + \tilde{\Psi}| W_{+} | \hat{\Psi} + \tilde{\Psi} \rangle
\ge \langle \hat{\Psi}| W_{-} | \hat{\Psi} \rangle.
\label{warunek1}
\end{eqnarray}

$\hat{V}$ is a product space. To see it, let's introduce the notations
\begin{displaymath}
\begin{array}{l}
 A_{\Psi} = \mathrm{Im} \mathrm{Tr}_2 | \Psi \rangle \langle \Psi | \\
 B_{\Psi} = \mathrm{Im} \mathrm{Tr}_1 | \Psi \rangle \langle \Psi |
\end{array}
\end{displaymath}
and write the subspace $\tilde{V}_{\Psi}$ using them:
\begin{displaymath}
 \tilde{V}_{\Psi} = \mathrm{span} \{ A_{\Psi} \otimes \mathbb{C}^{d_2} \ \cup \ \mathbb{C}^{d_1} \otimes B_{\Psi} \}.
\end{displaymath}
It's orthogonal complement is then equal:
\begin{displaymath}
\begin{array}{l}
 \tilde{V}_{\Psi}^{\perp} = (A_{\Psi} \otimes \mathbb{C}^{d_2})^{\perp} \ \cap \ (\mathbb{C}^{d_1} \otimes B_{\Psi})^{\perp} = \\ \\ A_{\Psi}^{\perp} \otimes \mathbb{C}^{d_2} \ \cap \ \mathbb{C}^{d_1} \otimes B_{\Psi}^{\perp} = A_{\Psi}^{\perp} \otimes B_{\Psi}^{\perp}
\end{array}
\end{displaymath}
The subspace $\hat{V}$ is an intersection of such subspaces, so it's also a product subspace:
\begin{displaymath}
\begin{array}{l}
 \hat{V} = \bigcap_{\Psi \in \mathcal{S}_{k} \cap V_{0}} \tilde{V}_{\Psi}^{\perp} = \bigcap_{\Psi \in \mathcal{S}_{k} \cap V_{0}} (A_{\Psi}^{\perp} \otimes B_{\Psi}^{\perp}) \\ \\ = (\bigcap_{\Psi \in \mathcal{S}_{k} \cap V_{0}} A_{\Psi}^{\perp}) \otimes (\bigcap_{\Psi \in \mathcal{S}_{k} \cap V_{0}} B_{\Psi}^{\perp}).
\end{array}
\end{displaymath}
Denote $\hat{V}_{1} = \bigcap_{\Psi \in \mathcal{S}_{k} \cap V_{0}} A_{\Psi}^{\perp}$ and $\hat{V}_{2} = \bigcap_{\Psi \in \mathcal{S}_{k} \cap V_{0}} B_{\Psi}^{\perp}$.

Because $\hat{V} = \hat{V}_{1} \otimes \hat{V}_{2}$ is a product subspace, after the apriopriate change of basis of $\mathbb{C}^{d_1}$ and $\mathbb{C}^{d_2}$ the matrix $\mathfrak{A}(\hat{\Psi})$ of an element $\hat{\Psi} \in \hat{V}$ is a matrix, which has zeros out of some subblock.
To see it, set such basis $\{ e \}_{i=1}^{d_1}$, $\{ f \}_{j=1}^{d_2}$ of  $\mathbb{C}^{d_1}$ and $\mathbb{C}^{d_2}$ respectively, that $\hat{V}_{1} = \mathrm{span} \{ e_{1}, \dots, e_{\mathrm{dim} \hat{V}_{1}} \}$ and $\hat{V}_{2} = \mathrm{span} \{ f_{1}, \dots, f_{\mathrm{dim} \hat{V}_{2}} \}$. Now the matrix $\mathfrak{A}(\hat{\Psi})$ of any element $\hat{\Psi} \in \hat{V}$ has zeros outside the left-upper corner of the size $\mathrm{dim} \hat{V}_{1} \times \mathrm{dim} \hat{V}_{2}$.

From now, we use such basis. The matrix  $\mathfrak{A}(\hat{\Psi})$ is therefore non-zero only in its left-upper  subblock of size $\mathrm{dim} \hat{V}_{1} \times \mathrm{dim} \hat{V}_{2}$, and the matrix  $\mathfrak{A}(\tilde{\Psi})$ is therefore non-zero only outside its left-upper  subblock of size $\mathrm{dim} \hat{V}_{1} \times \mathrm{dim} \hat{V}_{2}$. This means that a given vector $\Psi = \hat{\Psi} + \tilde{\Psi}$ is $k-$separable, only if the vector $\hat{\Psi}$ is $k-$separable
(If a matrix has the rank lower than $k$, then any its subblock has the rank lower than $k$).
Now for a given normalized vector $\hat{\Psi} \in \mathcal{S}_{k} \cap \hat{V}$ define the set
\begin{equation}
U_{\hat{\Psi}}=\{ \tilde{\Psi} \in \hat{V}^{\perp}: \hat{\Psi} + \tilde{\Psi} \in \mathcal{S}_{k} \}.
\label{U_Psi}
\end{equation}
Using this observation, one can rewrite the condition (\ref{warunek1}) in the following form
\begin{eqnarray}
\forall \hat{\Psi} \in \hat{V} \cap \mathcal{S}_{k}: ||\hat{\Psi}||=1 \ \forall \tilde{\Psi} \in U_{\hat{\Psi}} \nonumber \\
\langle \hat{\Psi} + \tilde{\Psi}| W_{+} | \hat{\Psi} + \tilde{\Psi} \rangle
\ge \langle \hat{\Psi}| W_{-} | \hat{\Psi} \rangle.
\label{warunek2}
\end{eqnarray}
The sets $U_{\hat{\Psi}}$ are:
\begin{itemize}
\item non-empty: Zero is always an element.
\item closed (in the metric topology of $\hat{V}^{\perp}$):

By the isomorphism $\mathfrak{A}$ defined by (\ref{mathfrakA}), the set $\mathfrak{A}(U_{\hat{\Psi}})$ is a set of matrices having their left-upper block of the size $\mathrm{dim} \hat{V}_{1} \times \mathrm{dim} \hat{V}_{2}$ fixed and the rank less or equal $k$ (direct reformulation of definition (\ref{U_Psi}) ). It is a set of common zeros of all minors of rank $k+1$ (the matrix entries from left-upper block we treat as fixed constants). The minors are polynomial functions, and hence continous, so the set $U_{\hat{\Psi}}$ is the inverse image of a closed set $\{0\}$, and therefore closed.
\end{itemize}

Now define a continous function $f_{\hat{\Psi}}: U_{\hat{\Psi}} \to \mathbb{R} \cup \{0\}$ by the left side of the inequality (\ref{warunek1})
\begin{equation}
f_{\hat{\Psi}} = \langle \hat{\Psi} + \tilde{\Psi}| W_{+} | \hat{\Psi} + \tilde{\Psi} \rangle.
\label{fPsi}
\end{equation}
By the second and the third assumption, we have:
\begin{displaymath}
\forall \Psi \in \mathcal{S}_{k} \quad \Psi \in V_{0} \oplus V_{-} \Rightarrow \Psi \in V_{0} \Rightarrow \Pi_{\hat{V}} \Psi = 0
\end{displaymath}
(here $\Pi_{\hat{V}}$ denotes an orthogonal projection). Now, by the contraposition rule
\begin{displaymath}
\forall \Psi \in \mathcal{S}_{k} \quad \Pi_{\hat{V}} \Psi \ne 0 \Rightarrow \Psi \not \in V_{0} \oplus V_{-} \Rightarrow \langle \Psi | W_{+}| \Psi \rangle > 0.
\end{displaymath}
Therefore on any $\tilde{\Psi} \in U_{\hat{\Psi}}$ the function $f_{\hat{\Psi}}$ takes stricty non-negative values. The image of the domain of $f_{\hat{\Psi}}$ is then a closed subset of $\mathbb{R}_{+}$. The infimum of the function $f_{\hat{\Psi}}$ on its domain is then stricty positive. Denote it by $c_{\hat{\Psi}}$.

Consider now the continous function $c: \hat{\Psi} \to c_{\hat{\Psi}}$, defined on the set $\{ \hat{\Psi}: \hat{\Psi} \in \hat{V} \ \land \ ||\Psi||=1\}$. The domain is compact, so the function $c$ is bounded from below, the lower bound $C$ is reached at some point of the domain and then is positive. This result allows to bound the left-hand side of the inequality in (\ref{warunek1}) from below as
\begin{displaymath}
\langle \hat{\Psi} + \tilde{\Psi}| W_{+} | \hat{\Psi} + \tilde{\Psi} \rangle \ge C
\end{displaymath}

On the same set $\{ \hat{\Psi}: \hat{\Psi} \in \hat{V} \cap \mathcal{S}_{k} \ \land \ ||\hat{\Psi}||=1\}$ we have also given a real-valued continous function by the right-hand side of the inequality (\ref{warunek1})
\begin{displaymath}
g(\hat{\Psi}) = \langle \hat{\Psi}| W_{-} | \hat{\Psi} \rangle.
\end{displaymath}
Again, because of compactness of the domain, this function is bounded from above. Denote its maximum by $G$.

Now by rescaling the positive part $W_{+}$ one obtains $C \ge G$. This ensures, that the inequality in (\ref{warunek}) is fulfilled for any vector of the Schmidt rank $k$, and then the observable is a $k$-SW.
$\square$

The next theorem will allow us to give some restrictions on the signature of $k-$Schmidt witnesses when the dimensions of the subsystems are given.

\begin{lem} The set $\mathcal{S}_{k} \subset \mathbb{C}^{d_1} \otimes \mathbb{C}^{d_2}$ is an affine variety of dimension $d_1 \times d_2 - (d_1 - k) \times (d_2 - k)$.
\end{lem}

\textbf{Proof:}
By the isomorphisms $\mathfrak{A}$ defined by (\ref{mathfrakA}) we can treat the set $\mathcal{S}_{k}$ as a set of $d_1 \times d_2$ matrices of rank $k$. It is an affine variety generated by the ideal of all minors of rank $k+1$. There are $\left( \begin{array}{c} d_1 \\ k+1 \end{array} \right) \times \left( \begin{array}{c} d_2 \\ k+1 \end{array} \right)$ such a minors, but locally only $(d_1-k) \times (d_2-k)$ of them are indepentent.

For any regular point of this variety $A$ (a matrix of the rank equal $k$) we can find basis in $\mathbb{C}^{d_1}$ and $\mathbb{C}^{d_2}$, respectively $\{ f \}_{i=1}^{d_1}$ and $\{ e \}_{i=1}^{d_2}$ such that $A = \sum_{i=1}^{k} f_{i}^{T} e_{i}$. In this basis a minor built from the first $k$ columns and the first $k$ rows is non-zero. There exists an open neighbourhood of this regular point (in the space $\mathbb{M}_{\mathbb{C}}(d_1 \times d_2)$), such that any matrix in this neighbourhood has its first $k$ columns and first $k$ rows linearly independent. To check whether a given matrix from this neighbourhood has rank less or equal $k$, one has to check $(d_1 - k) \times (d_2 - k)$ independent conditions
\begin{displaymath}
\mathrm{det} \left[ \begin{array}{ccc|c} b_{11} & \dots & b_{1k} & b_{1i} \\ \vdots & \ddots & \vdots & \vdots \\ b_{k1} & \dots & b_{kk} & b_{ki} \\ \hline b_{j1} & \dots & b_{jk} & b_{ji} 
\end{array} \right] = 0.
\end{displaymath}
(in the new basis). Dimension of the tangent space in a regular point, and hence the dimension of the variety is then equal $d_1 \times d_2 - (d_1 - k) \times (d_2 - k)$.
$\square$

For detailed discusion and proofs of the facts about manifolds of matrices of a fixed rank see \cite{KusMarmo} or \cite{DifGeomCh}

\begin{twr} Any subspace $V \in \mathbb{C}^{d_1} \otimes \mathbb{C}^{d_2}$ of dimension $\mathrm{dim}V > (d_1 - k) \times (d_2 - k)$ contains a $k-$separable vector.
\label{max-k-sep}
\end{twr}

\textbf{Proof:}
Because the variety $\mathcal{S}_{k}$ is defined by a zero of an ideal of homogenous polynomials, one can consider its projectivisation of the dimension $d_1 \times d_2 - (d_1 - k) \times (d_2 - k) - 1$. It is a subvariety of $\mathbb{C}P^{d_1 \times d_2 -1}$. Consider now another subvariety of $\mathbb{C}P^{d_1 \times d_2 -1}$ - a projectivisation of linear subspace $V \subset \mathbb{C}^{d_1} \otimes \mathbb{C}^{d_2}$. Its dimension is $\mathrm{dim}V-1$. Projective Dimension Theorem \cite{Hartshorne} says that if the sum of dimensions of two subvarieties of $\mathbb{C}P^{N}$ is greater or equal $N$, then their intersection is a nonempty set. Using this for the above subvarieties, we find than if
\begin{displaymath}
d_1 \times d_2 - (d_1 - k) \times (d_2 - k) - 1 + \mathrm{dim}V-1 \ge d_1 \times d_2 - 1,
\end{displaymath}
then subspace $V$ consists of some $k-$separable vectors.
$\square$

The theorem used is a generalization of the well-known fact, that two projective lines on a projective plane always have an intersection. To show that there is no stricter restriction on this dimension, I give an example of basis (non-orthogonal), which spans the $(d_1-k) \times (d_2-k)$-dimensional subspace, which does not contain any $k-$separable vector.

\begin{twr}
The set $\{ \sum_{i=1}^{k} e_{m+i} \otimes f_{n+i} \}_{m \le d_1-k, n \le d_2-k}$ spans the subspace $V_{max}^{k}$ which does not contain any non-zero $k-$separable vector.
\end{twr}

\textbf{Proof:} 
Again using the isomorphism $\mathfrak{A}$ defined by (\ref{mathfrakA}), we will prove that any non-zero matrix in the subspace $V_{max}^{k} \subset \mathbb{M}_{\mathbb{C}}(d_1 \times d_2)$ spanned by the set of matrices $\{ \sum_{i=1}^{k} e_{m+i}^{T} f_{n+i} \}_{m \le d_1-k, n \le d_2-k}$ has rank greater than $k$.

We call any subset of matrix elements with a constant difference between indices a \textit{diagonal} of a matrix (in general rectangular). This difference will be called the \textit{number} of a diagonal. The number of a diagonal varies between $1-d_1$ and $d_2 -1$.

The proof will be carried out by induction with respect to the number of a diagonal. Any matrix in $V_{max}^{k}$ has the first $k$ and the last $k$ of diagonals equal zero (any diagonal which has less than $k$ elements is equal zero). Now assume, that there exists a matrix of the rank $k$ in $V_{max}^{k}$.

We will prove that if all diagonals of numbers less than $p$ are zero, then also the $p$-th diagonal is zero.

Any minor of rank $k+1$ is equal zero, in particular the minors whose diagonal (in standard sense) is created from elements of the considered $p$-th diagonal. Because we assume that all diagonals with their numbers less than $p$ are equal zero, such minors are determinats of upper-triangular matrices, and therefore the products of $k+1$ elements from the $p$-th diagonal. If all such minors are equal zero, then all products of $k+1$ elements from the considered diagonal are equal zero. All basis elements which give a non-zero contribution to $p-$th diagonal are $v_{m} = \sum_{i=1}^{k} e_{m+i} \otimes f_{n+i}$, where $n-m=p$. In the same way let us number the elements of diagonal $b_{m}$. Teh relation between elements of diagonal and the coefficients of combination of $v_{m}$ is given by the following system of linear equations:
\begin{equation}
\left[ \begin{array}{cccc}
1      & 0      & \dots  & 0      \\
\vdots & \ddots & \ddots & \vdots \\
1      &        & \ddots & 0      \\
0      & \ddots &        & 1      \\
\vdots & \ddots & \ddots & \vdots \\
0      & \dots  & 0      & 1      \\
\end{array} \right]
\left[ \begin{array}{c} \alpha_{1} \\ \vdots \\ \alpha_{k+1} \\ \\ \vdots \\ \
\end{array} \right] =
\left[ \begin{array}{c} 
b_{1} \\ \vdots \\ b_{k+1} \\ \\ \vdots \\ \
\end{array} \right]
\label{p_diag}
\end{equation}
Because any product of $k+1$ elements from the considered diagonal is equal zero, at most $k$ equations can have the right-hand side different from $0$. Removing any $k$ rows from the matrix in (\ref{p_diag}), we have a nonsingular square matrix. Because we already removed all equations with non-zero right-hand side, the system (\ref{p_diag}) has only zero solution. By induction, we then zero all diagonals, so the only matrix in $V_{max}^{k}$ with rank less than $k$ is the matrix with all coefficients equal $0$.
$\square$

\section{Propositions and examples}

\begin{pr}
Any eigenvector of a $k$-SW related to a negative eigenvalue is not $k-$separable
\end{pr}

\textbf{Proof:} Use the second condition in the theorem \ref{main_th_need}.$\square$

The theorem \ref{max-k-sep} leads to proposition:

\begin{pr}
For $k$-SW, $1 \le \mathrm{dim} V_{-} \le (d_1 - k) \times (d_2 - k)$. 
\label{dimV-}
\end{pr}

If in $V_{0}$ there are no $k-$separable vectors, then the third condition in the theorem \ref{main_th_need} is fulfilled, and then we have the following:

\begin{pr}
If $\mathrm{ker} W \cap \mathcal{S}_{k} = \emptyset$ and $W$ is an $k$-SW, then
\begin{displaymath}
\mathrm{dim} V_{+} \ge k(d_1 + d_2) - k^{2}
\end{displaymath}
\label{dimV+}
\end{pr}

\textbf{Proof:} By the second condition in Theorem \ref{main_th_need}, if there is a $k-$separable vector in $V_{0} \oplus V_{-}$ then it belongs to $V_{0}$. Because by the assumption no $k-$separable vectors are in $V_{0}$, no $k-$separable vectors are in $V_{0} \oplus V_{-}$, and therefore the dimension of $V_{0} \oplus V_{-}$ is bounded from above by $(d_1 - k) \times (d_2 - k)$. Because $\mathrm{dim} V_{+} = d_1 d_2 - \mathrm{dim} (V_{-} \oplus V_{0})$, we get the above inequality. $\square$

It is not possible to establish any non-trivial upper bound for positive subspace of an arbitrary $k$-Schmidt witness, because we can always get a new $k$-SW by adding a positive observable, with its support contained in the kernel of the witness. An interesting question would be, whether there exist such bounds for optimal witnesses, i.e. such witnesses, from which no positive observable cannot be subtracted without destroying the property of being $k$-Schmidt witness.

We can construct in practice examples of subspaces, which do not contain any separable vector, by means of \textit{unextendible product basis} (denoted further as UPB, see \cite{UPB_99}, \cite{UPB_98}), as the orthogonal complement of a subspace spanned by UPB. Such a subspace by the definion of UPB contains no separable vectors. The maximal number $n$ of vectors in UPB in $\mathbb{C}^{d_1} \otimes \mathbb{C}^{d_2}$ is bounded from below by the number
\begin{displaymath}
n \ge \sum_{i} (d_i - 1) + 1
\end{displaymath}
(see Lemma 1 in \cite{UPB_99}). Subtracting it from the dimension of the space, we get an upper bound for the dimension of the constructed subspace, which does not contain any separable vectors
\begin{displaymath}
\dim V \le (d_1 - 1) \times (d_2 - 1).
\end{displaymath}
Then, for subspaces constructed by means of UPB, Proposition \ref{dimV-} recovers the previous results.

Now for a subspace $V$ with no non-negative separable vectors, Theorem \ref{suff} assures us that an observable:
\begin{displaymath} 
W = I - (1+\epsilon) P_{V}
\end{displaymath}
is an EW for small enough positive $\epsilon$. ($P_{V}$ denotes a projection onto subspace $V$). For subspaces constructed by means of UPB, such a construction was made in \cite{Terh2}. Moreover, the nondecomposability of such W was proven.

\begin{ex}
Any EW for 2 qubits has exactly one negative eigenvalue and exactly three positive eigenvalues.
\end{ex}

\textbf{Proof:}
At least one negative eigenvalue is needed, to ensure that the witness detects anything (first condition in Definition \ref{k_ent_wit}). The eigenvector related to this eigenvalue is of Schmidt rank two. It will be denoted as $\Psi^{-}$. The negative space cannot consist of any separable vector, so Proposition \ref{dimV-} bounds its dimension by 1.

No vector of Schmidt rank greater than one can be in the kernel. If there were such a vector $\Psi$, then by Nullstellensatz there exists such $\alpha:\beta$, that
\begin{displaymath}
\mathrm{rank} (\alpha \mathfrak{A}(\Psi^{-}) + \beta \mathfrak{A}(\Psi)) = 1,
\end{displaymath}
but then $\beta \ne 0$, so there would be a separable vector $\alpha \Psi + \beta \Psi^{-} \in V_{0} \oplus V_{-}$, but $\alpha \Psi + \beta \Psi^{-} \in V_{0} \oplus V_{-} \not \in V_{0}$, which would be in contradiction with the second condition in Theorem \ref{main_th_need}. Thus there can be only separable vectors in the kernel.

Write such a vector in its Schmidt decomposition
\begin{displaymath}
\mathfrak{A}(\Psi^{0}) = \left[ \begin{array}{cc} 1 & 0 \\ 0 & 0 \end{array} \right].
\end{displaymath}
The vector $\Psi^{-}$ is of Schmidt rank greater than one and orthogonal to $\Psi^{0}$ and can be written as
\begin{displaymath}
\mathfrak{A}(\Psi^{-}) = \left[ \begin{array}{cc} 0 & a \\ b & c \end{array} \right],
\end{displaymath}
where $ab \ne 0$.
Now let us identify all separable vectors in $V_{0} \oplus V_{-}$
\begin{displaymath}
\mathrm{rank} (\alpha \mathfrak{A}(\Psi^{-}) + \beta \mathfrak{A}(\Psi^{0})) = 1 \quad \iff \quad \beta (\alpha c + \beta ab) = 0.
\end{displaymath}
If $c \ne 0$, then there exists a separable vector, which is in $V_{0} \oplus V_{-}$, but not in $V_{0}$, which is in contradiction with the second condition in Theorem \ref{main_th_need}. Thus $c=0$.

If there is any other vector in the kernel, let us say $\Psi^{0}_{1}$, then by the condition that no entangled vector is in $V_{0}$, the subspace $V_{0}$ consists of only separable vectors. We then conclude that $\mathfrak{A}(\Psi^{0}_{1})$ has the second column or the second row equal $0$, but no such a vector (other than $\Psi^{0}$) is orthogonal to $\Psi^{-}$. Therefore there is at most one vector in the kernel (up to scaling by nonzero scalar) and it is separable.

Let us identify now the subspace $\tilde{V}_{\Psi^{0}}$, which will be needed to check the third condition in Theorem \ref{main_th_need}. It is the subspace of all vectors $\Psi$, for which $[\mathfrak{A}(\Psi)]_{22} = 0$. Observe that $\Psi^{-} \in \tilde{V}_{\Psi^{0}}$, which is in contradiction with the third condition in Theorem \ref{main_th_need}. No non-zero vector thus can be in the kernel.$\square$

Because we know that any EW in $\mathbb{C}^{2} \otimes \mathbb{C}^{2}$ is decomposable \cite{decomp}, we conclude that the partial transposition of an entangled positive matrix in $\mathbb{C}^{2} \otimes \mathbb{C}^{2}$ has exactly one negative and $3$ positive eigenvalues.
For another proof of this fact, see \cite{Aug}

\section{Conditions on eigenvalues of entanglement witnesses}

Theorem \ref{suff} assures, that given two semipositive observables $W_{\pm}$ which have othogonal supports, when some conditions on the support of $W_{-}$ are fulfilled, one can construct a witness $\lambda W_{+} - W_{-}$ for sufficiently large alpha. There is still a question of how big $\lambda$ should be. We restrict ourselves in this subsection to entanglement witnesses, i.e. we fix $k=1$. In some cases, for $k=1$ parameter $\lambda$ can be calculated or at least it acquires some new interpretation.

In the beginning, take $W_{-} = P_{V_{-}}$ and $W_{+} = I - P_{V_{-}}$. Because $V_{0} = (V_{+} \oplus V_{-})^{\perp} = \{ 0 \}$, the third condition of Theorem \ref{suff} is fulfilled in a trivial way. The second one in this situation reads $V_{-} \cap \mathcal{S}_{k} = \emptyset$. Let us take therefore a subspace $V_{-}$ which does not contain any vectors of Schmidt rank one.

Up to normalization, the witness $\lambda W_{+} - W_{-}$ is equal to
\begin{displaymath}
 W = \epsilon I - P_{V_{-}}, \quad \epsilon = \frac{\lambda}{\lambda+1}.
\end{displaymath}
Because the observable is an entanglement witness, it has positive eigenvalues when restricted to separable states. For any normalized $\Psi \in \mathcal{S}_{1}$ we have therefore
\begin{displaymath}
 \epsilon \ge \langle \Psi | P_{V_{-}} | \Psi \rangle = || P_{V_{-}} \Psi ||^{2} = ( \sup \{ | \langle \Phi | \Psi \rangle | : \ \Phi \in V_{-} \ \land \ || \Phi || = 1 \} )^{2}.
\end{displaymath}
The smallest $\epsilon$ which fulfils the above condition is a supremum of the right hand side with respect to all normalized $\Psi = \phi \otimes \chi$,
\begin{displaymath}
 \epsilon_{min} = (\sup \{ | \langle \Phi|\phi \otimes \chi \rangle |: \ ||\phi|| = 1 \ \land \ ||\chi||=1 \ \land \ ||\Phi||=1 \ \land \ \Phi \in V_{-}\} )^{2}.
\end{displaymath}
It is easy to observe, that $\langle \Phi|\phi \otimes \chi \rangle = \phi^{T} \mathfrak{A}(\Phi) \chi$. Using it, we find the final formula for $\epsilon$:
\begin{eqnarray*}
 \epsilon_{min} = (\sup \{ | \phi^{T} \mathfrak{A}(\Phi) \chi |: \ ||\phi|| = 1 \ \land \ ||\chi||=1 \ \land \  ||\Phi||=1  \ \land \ \Phi \in V_{-}\} )^{2} \nonumber \\ \\
= (\sup \{ ||\mathfrak{A}(\Phi)||: ||\Phi||=1  \ \land \ \Phi \in V_{-} \})^{2} \nonumber \\ \\
= \sup \{ ||\mathfrak{A}(\Phi)||^{2}: ||\Phi||=1  \ \land \ \Phi \in V_{-} \}.
\end{eqnarray*}
The supremum norm of $\mathfrak{A}(\Psi)$ is the maximal Schmidt coefficient of the vector $\Psi$. Translating by the isomorphism $\mathfrak{A}$ the subspace $V_{-}$ into the corresponding subspace $\mathfrak{A}(V_{-})$ in the space of matrices of coefficients, we are looking for the supremum of the sup norm on the unit sphere in this subspace. The quantity $\epsilon_{min}$ can be therefore interpreted as the \textit{supremum norm of the subspace of matrices}. We will therefore denote it further as $\epsilon_{min} = ||\mathfrak{A}(V_{-})||_{sup}^{2}$. When the subspace $V_{-}$ is one-dimensional and spanned by a vector $\Psi$, $\epsilon_{min}$ can be calculated as $||\mathfrak{A}(\Psi)||_{sup}^{2}$.
We can use this special class of witnesses to find the conditions for the eigenvalues of entanglement witnesses. One can estimate any witness $W = W_{+} - W_{-}$ from above as
\begin{eqnarray}
 W_{+} - W_{-} \le \lambda_{+}^{max} P_{V_{+}} - \lambda_{-}^{min} P_{V_{-}} \le \lambda_{+}^{max} P_{V_{+} \oplus V_{0}} - \lambda_{-}^{min} P_{V_{-}}
\nonumber \\
\nonumber \\
= \lambda_{+}^{max} I - (\lambda_{+}^{max} + \lambda_{-}^{min}) P_{V_{-}},
\end{eqnarray}
and from below as
\begin{eqnarray}
W_{+} - W_{-} \ge \lambda^{+}_{min} P_{V_{+}} - \lambda^{-}_{max} P_{V_{-}}
\ge \lambda_{+}^{min} P_{V_{+}} - \lambda_{-}^{max} P_{V_{-} \oplus V_{0}}
\nonumber \\
\nonumber \\
= \lambda_{+}^{min} I - (\lambda_{+}^{min} + \lambda_{-}^{max}) P_{V_{-} \oplus V_{0}}.
\end{eqnarray}

The upper bound is EW iff
\begin{equation}
 \frac{\lambda_{+}^{max}}{\lambda_{+}^{max}+\lambda_{-}^{min}} \le ||\mathfrak{A}(V_{-})||_{sup}^{2}.
\label{suff_W}
\end{equation}
We conclude therefore, that (\ref{suff_W}) is a necessary condition for $W$ to be entanglement witness.

The lower bound is EW iff
\begin{equation}
 \frac{\lambda_{+}^{min}}{\lambda_{+}^{min}+\lambda_{-}^{max}} \le ||\mathfrak{A}(V_{-} \oplus V_{0})||_{sup}^{2}
\label{necess_W}.
\end{equation}
We conclude now, that (\ref{necess_W}) is a sufficient condition for $W$ to be an entanglement witness. Observe, that this sufficient condition can be fulfilled iff there are no vectors of Schmidt rank $1$ in the kernel.

When the negative subspace of $W$ is one-dimensional and the kernel is trivial, then one gets the sufficient condition
\begin{displaymath}
 \frac{\lambda_{+}^{min}}{\lambda_{+}^{min}+\lambda_{-}} \le ||\mathfrak{A}(\Psi)||_{sup}^{2}
\end{displaymath}
One can find this condition for related positive maps in \cite{BenFlor}.

It rests to show how Theorem \ref{suff} works when there exist vectors of Schmidt rank $1$ in the kernel. In this example, consider the space $\mathbb{C}^{3} \otimes \mathbb{C}^{4}$. Denote by $\{ e_{i} \}_{i=1}^{3}$ the basis of $\mathbb{C}^{3}$ and by $\{ f_{i} \}_{i=1}^{4}$ the basis of $\mathbb{C}^{4}$ Let us span the kernel by two vectors: $e_{1} \otimes f_{1}$ and $e_{1} \otimes f_{2}$. The space $\hat{V}$ is now
\begin{displaymath}
\hat{V} = \mathrm{span} \{ e_{2}, e_{3} \} \otimes \mathrm{span} \{ f_{3}, f_{4} \}.
\end{displaymath}
The subspace $V_{-}$ can be now embeded in a two-qubit space. It is therefore spanned by one normalized vector of Schmidt rank two. Denote it by $\psi_{-}$. Now we choose arbitrary orthonormal basis of $V_{-}^{\perp} \cap \mathrm{span} \{ e_{2}, e_{3} \} \otimes \mathrm{span} \{ f_{3}, f_{4} \}$. Denote its elements by $k_1$, $k_3$ and $k_3$. Finaly, we complete the set $e_{1} \otimes f_{1}, e_{1} \otimes f_{2}, \psi_{-}, k_1, k_2, k_3 \}$ to the orthonormal basis of $\mathbb{C}^{3} \otimes \mathbb{C}^{4}$ by six vectors $l_1, ..., l_6$. It will be the basis of our witness.

Now Theorem \ref{suff} assures, that fixing the negative eigenvalue, say $-1$, and choosing large enough positive eigenvalues, one can get an entanglement witness. To do it, let us fix the eigenvalues related to the vectors $k_1$, $k_3$ and $k_3$ to be equal $\epsilon / (1-\epsilon)$, where $\epsilon = ||\mathfrak{A}(\psi_{-})||_{sup}^{2}$. Suppose, that the rest of eigenvalues are equal zero --- we have two-qubit witness whose domain is embeded in $\mathbb{C}^{3} \otimes \mathbb{C}^{4}$. We can leave the eigenvalues related to the eigenvectors $l_{1}, \dots, l_{6}$ unchanged, or increase them to some positive values --- it only destroys the optimality of the witness.

\section{Translation to some properties of positivity-preserving mappings between the matrix algebras}

The set of linear maps between the matrix algebras $\mathcal{B}(\mathbb{C}^{d_1})$ and $(\mathcal{B}(\mathbb{C}^{d_2}))$ and the set of bilinear forms on $\mathbb{C}^{d_1} \otimes \mathbb{C}^{d_2}$ is isomorphic by the well-known Jamio\l kowski isomorphisms \cite{Jam} $J: \mathcal{B}(\mathbb{C}^{d_1} \otimes \mathbb{C}^{d_2}) \to \mathcal{L}(\mathcal{B}(\mathbb{C}^{d_1}), \mathcal{B}(\mathbb{C}^{d_2}))$,
\begin{equation}
W_{\Lambda} = J(\Lambda) = [I \otimes \Lambda] | \Psi^{+} \rangle \langle \Psi^{+} |,
\label{jam}
\end{equation}
where $\mathfrak{A}(\Psi^{+}) = 1/d_1 \ \mathrm{Id}_{d_1}$ (a projector onto $\Psi^{+}$ is the maximally entangled state).

The subset of linear maps, which preserve the hermiticity of a matrix, is isomorphic by (\ref{jam}) to the subset of Hermitian bilinear forms. Any Hermitian matrix has spectral decomposition
\begin{equation}
W = \sum_{i=1}^{p} \lambda^{+}_{i} |\Psi^{+}_{i} \rangle \langle \Psi^{+}_{i} | - \sum_{i=1}^{q} |\lambda^{-}_{i}| |\Psi^{-}_{i} \rangle \langle \Psi^{-}_{i} |,
\label{herm-obs}
\end{equation}
where $(p,q)$ is the signature of $W$. Such a matrix is related by the isomorphism (\ref{jam}) to a hermiticity-preserving linear map:
\begin{equation}
\Lambda{\rho} = \sum_{i=1}^{p} A_{i} \rho A_{i}^{\dagger} - \sum_{i=1}^{q} B_{i} \rho B_{i}^{\dagger},
\label{herm-pres}
\end{equation}
where $A_{i} = \sqrt{\lambda^{+}_{i}} \mathfrak{A}(\Psi^{+}_{i})$ and $B_{i} = \sqrt{|\lambda^{-}_{i}|} \mathfrak{A}(\Psi^{-}_{i})$. For more details and other facts about this correspondence see \cite{Ali} and the references therein.

The matrices $A_{i}$ and $B_{i}$ are orthogonal with respect to the Hilbert-Schmidt inner product. The observable $W$ admits many decompositions into linear combinations of projectors, but the decomposition into a combination of orthogonal projectors is unique up to degeneracy of eigenvalues. Similarly, one hermiticity-preserving map $\Lambda$ admits many decompositions into the form (\ref{herm-pres}) (such a form is called the \textit{Kraus-Choi form} of hermiticity preserving map due to Kraus-Choi representation theorem \cite{Kraus-Choi}), but if we assume the ortogonality of matrices $A_{i}$ and $B_{i}$, then the decomposition becomes unique.

In the set of maps, which preserve the hermiticity we are allowed to define a subset of maps which preserve positivity. Such maps are called \textit{positive maps}. We can generalize the definition of a positive map to $k$-positive map such that $\Lambda$ is $k$- positive when the map $Id_{k} \otimes \Lambda$ is positive. The isomorphism (\ref{jam}) relates a $k$-positive map to a $k$-Schmidt witness. The isomorphism (\ref{jam}) allows to reformulate the propositions about properties of $k$-SWs to propositions describing properties of $k-$positive maps. Observe that if matrices $A_{i}$, $B_{i}$ in (\ref{herm-pres}) are not orthogonal, but still lineary independent, then $(p,q)$ remains unchanged.

Cosider now a hermiticity-preserving map in its Kraus-Choi form:
\begin{equation}
\Lambda{\rho} = \sum_{i=1}^{p} A_{i} \rho A_{i}^{\dagger} - \sum_{i=1}^{q} B_{i} \rho B_{i}^{\dagger}.
\label{K-C}
\end{equation}
Assume that matrices $A_{i}, B_{i}$ are linearly independent. We have then the following:

\begin{pr}
If $\Lambda$ is a $k$-positive map, then $0 \le q \le (d_1 - k) \times (d_2 - k)$.
\end{pr}

\begin{pr}
If $\Lambda$ is a $k$-positive map and does not map any state of rank less or equal $k$ to zero, then $p \ge d_1 d_2 - (d_1-k) \times (d_2-k)$.
\end{pr}

\section{Conclusions}

The necessary and sufficient condition for $k$-Schmidt witness has been presented and proved. Moreover, in the system of two qubits, the necessary condition fully determines the spectral type of any entanglement witness. The neccesary condition provides also an upper bound for the dimension of negative subspace of a $k$-Schmidt witness in arbitrary dimensions, which can be translated to analogous properties of $k-$positive maps. Also the sufficient condition for a $k$-Schmidt witness (up to rescaling of the positive part of an observable) has been proved.

\ack
This work was partially supported by the
Polish Ministry of Science and Higher Education Grant No
3004/B/H03/2007/33.
I would like to thank prof. Ma\'{c}kowiak and dr Michalski for language corrections and to prof. Zwara for fruitfull discussions about algebraic geometry.


\vspace{0.1cm}

\section{References}

\begin{thebibliography}{99}

\bibitem{Cox} H.S.M. Coxeter \textit{Projective geometry} (Springer, 2nd ed. 1974)

\bibitem{Hartshorne} R. Hartshorne \textit{Algebraic Geometry} (Graduate Texts in Mathematics 52, 1997)

\bibitem{Wern} R. F. Werner \textit{Quantum states with Einstein-Podolsky-Rosen correlations admitting a hidden-variable model} Phys. Rev. A \textbf{40} 4277 (1989)

\bibitem{Terh1} B. M. Terhal \textit{Bell Inequalities and the Separability Criterion} Phys. Lett. A \textbf{271} 319, eprint quant-ph/9911057 (2000)

\bibitem{TerhHor} B.M. Terhal, P. Horodecki \textit{A Schmidt number for density matrices} Phys. Rev. A \textbf{61} 040301(R) (2000), eprint quant-ph/9911117

\bibitem{Sanp} A. Sanpera, D. Bruss, M. Lewenstein \textit{Schmidt number witnesses and bound entanglement} Phys. Rev. A \textbf{63} 050301 (2001), eprint quant-ph/0009109 



\bibitem{KusMarmo} J. Grabowski, M. Ku\'s, G. Marmo \textit{Geometry of quantum systems: density states and entanglement} J.Phys. A \textbf{38} (2005) 10217-10244, eprint math-ph/0507045

\bibitem{DifGeomCh} Zuhuan Yua, Xianqing Jost-Lib, Qingzhong Lia, Jintao Lva, and Shao-Ming Feia \textit{Differential Geometry of Bipartite Quantum States} eprint quant-ph/0711.1211v1 (2007)

\bibitem{UPB_99} D. P. DiVincenzo, T. Mor, P. W. Shor, J. A. Smolin, B. M. Terhal \textit{Unextendible Product Bases, Uncompletable Product Bases and Bound Entanglement} Comm. Math. Phys. 238, pp. 379-410 (2003), quant-ph//9908070

\bibitem{UPB_98} C. H. Bennett, D. P. DiVincenzo, T. Mor, P. W. Shor, J. A. Smolin, B. M. Terhal \textit{Unextendible Product Bases and Bound Entanglement} Phys. Rev. Lett. \textbf{82}, 5385 - 5388 (1999), quant-ph//9808030

\bibitem{Terh2} B. M. Terhal \textit{A Family of Indecomposable Positive Linear Maps based on Entangled Quantum States} eprint quant-ph/9810091 v3 (2000)

\bibitem{decomp} M. Lewenstein, B. Kraus, J.I. Cirac, P. Horodecki \textit{Optimization of entanglement witnesses} Phys. Rev. A \textbf{62} 052310 (2000), eprint quant-ph/0005014

\bibitem{Aug} R. Augusiak, P. Horodecki, M. Demianowicz \textit{Universal observable detecting all two-qubit entanglement and determinant based separability tests} (2006), eprint quant-ph/0604109

\bibitem{BenFlor} F.Benatti, R.Floreanini, M.Piani \textit{Non-decomposable Quantum Dynamical Semigroups and Bound Entangled States} quant-ph//0411095 (2004)

\bibitem{Jam} A. Jamio\l kowski Rep. Math. Phys. \textbf{3} 275 (1972)

\bibitem{Ali} K. S. Ranade, M. Ali \textit{The Jamiolkowski isomorphism and a conceptionally simple proof for the correspondence between vectors having Schmidt number k and k-positive maps} (2007), eprint quant-ph/0702255

\bibitem{Kraus-Choi} K. Kraus \textit{States, Effects and Operations: Fundamental Notions of Quantum Theory}
Springer-Verlag (1983)

\end{thebibliography}
\end{document}